\documentclass[11pt]{article}
\usepackage{epsfig}
\newcommand{\nn}{\nonumber}

\newcommand{\be}{\begin{equation}}
\newcommand{\ee}{\end{equation}}
\newcommand{\bea}{\begin{eqnarray}}
\newcommand{\eea}{\end{eqnarray}}

\newcommand{\gev}{ {\rm GeV} }

\def\beq{\begin{equation}}
\def\eeq{\end{equation}}

\newcommand{\tetaot}{\mbox{$\theta_{13}$}}

\newcommand{\deltt}{\mbox{$\Delta_{23}$}}
\newcommand{\delot}{\mbox{$\Delta_{13}$}}

\begin{document}
%
%
\thispagestyle{empty}
\begin{flushright}
{CERN-TH/2001-081}\\
{FTUAM-01-05}\\
{IFT-UAM/CSIC-01-10} \\
{FTUV-010323}\\
{IFIC/01-10}
\end{flushright}
\vspace*{1cm}
\begin{center}
{\Large{\bf On the measurement of leptonic CP violation} }\\
\vspace{.5cm}
J. Burguet Castell$^{\rm a,}$, M.B. Gavela$^{\rm b,}$\footnote{gavela@delta.ft.uam.es}, J.J. G\'omez Cadenas$^{\rm a,c}$\footnote{gomez@hal.ific.uv.es},
P. Hern\'andez$^{\rm c,}$\footnote{pilar.hernandez@cern.ch. On leave 
from Dept. de F\'{\i}sica Te\'orica, Universidad de Valencia.} 
O. Mena$^{\rm b,}$\footnote{mena@delta.ft.uam.es}
 
\vspace*{1cm}
$^{\rm a}$ Dept. de F\'{\i}sica At\'omica y Nuclear and IFIC, Universidad de Valencia, Spain \\
$^{\rm b}$ Dept. de F\'{\i}sica Te\'orica, Univ. Aut\'onoma de
Madrid, 28049 Spain \\
$^{\rm c}$ CERN, 1211 Geneva 23, Switzerland

\vspace{.3cm}

%
%
%






%
\begin{abstract}
\noindent

We show that the simultaneous determination of
the leptonic CP-odd phase $\delta$ and the angle $\theta_{13}$ from
 the subleading transitions $\nu_e\rightarrow\nu_\mu$ and
${\bar\nu}_e\rightarrow{\bar\nu}_\mu$ results generically, at fixed neutrino
energy and baseline, in two degenerate solutions. 
In light of this, we refine a previous analysis of
the sensitivity to leptonic CP violation at a neutrino factory, in the
LMA-MSW scenario, by exploring the full range of $\delta$ and $\theta_{13}$.
 Furthermore, we take into account the expected
uncertainties on the solar and atmospheric oscillation parameters and
in the average Earth matter density along the neutrino path.  An
intermediate baseline of $O(3000)$ km is still the best option to
tackle CP violation, although a combination of two baselines turns out
to be very important in resolving degeneracies.

\end{abstract}

\end{center}
%
%

\pagestyle{plain} 
\setcounter{page}{1}

%
%
\section{Introduction}
%

Solar parameters in the range of the large mixing angle MSW 
solution (LMA-MSW)\cite{wolf, solar}
are expected to affect sizeably the neutrino 
oscillation\cite{ponte} probabilities at terrestial 
distances. The discovery of CP-violation in the lepton 
sector might then be at reach \cite{cpold}-\cite{lipari}. 

 Consider a neutrino factory from a muon storage ring \cite{history, geer} 
with muon energies of 
 some dozens of GeV.
 In \cite{golden} a detailed study of the potential of such a neutrino 
factory in determining the leptonic CP-violating phase, $\delta$, and the 
angle $\theta_{13}$ was performed, 
within the LMA-MSW. 
Just the range $0 < \delta < 90^\circ$ and $\theta_{13} > 1^\circ$ 
was analyzed, though. 
It has been argued that the CP phase can be only determined up to a 
sign \cite{lipari}. In order to clarify this issue the full range 
$-180^\circ< \delta < 180^\circ$ has to be explored. Besides, it is 
important to understand what is the lower value of $\theta_{13}$ at which 
the sensitivity to CP violation is lost. 

A second limitation of the analysis performed in \cite{golden} is that the 
atmospheric parameters, $\theta_{23}$ and $\Delta m^2_{23}$, as well as the
solar ones, $\theta_{12}$ and $\Delta m^2_{12}$, were assumed to be known.  
Neither an error estimate on the average Earth matter density was included. 
It has been recently pointed out \cite{sato} that the errors in these 
parameters might modify drastically the conclusions reached in \cite{golden}.

For the atmospheric parameters, the main source of new information
will come from proposed long baseline accelarator experiments (KEK,
Minos, Opera)\cite{lbl} while for the solar ones lying in the LMA-MSW
range, mostly from Kamland\cite{kamproposal}.  Minos is expected to
measure $\sin^2 2 \theta_{23}$ and $\Delta m^2_{23}$ at the $10\%$
level \cite{minos}, in the range allowed by SuperKamiokande \cite{sk}.
In the neutrino factory this uncertainty is expected to be improved to
about $1\%$ from muon disappearance
measurements\cite{barger,concha,bcr}. Concerning the solar parameters
in the LMA-MSW regime, $\sin^2 2 \theta_{12}$ and $\Delta m_{12}^2$,
Kamland will be able to determine them to better than $10\%$ for
$\Delta m^2_{12} \geq 10^{-5}$ eV$^2$ and $\sin^2 2 \theta_{12} > 0.7$
\cite{kamland}, well before the time of the neutrino factory.

In this work we study, by means of an approximate analytical formula,
 the existence of degenerate solutions for the parameters ($\tetaot,\delta$): 
solutions that
 give the same probabilities for the $\nu_e\nu_\mu$ and 
${\bar\nu}_e{\bar \nu}_\mu$ transitions than the one chosen by nature. 
We complete the analysis of \cite{golden} 
by considering the full range of $\theta_{13}$ and $\delta$, 
which reveals the existence 
of these degeneracies,  and include  
the errors with which the atmospheric and solar parameters will be 
known at the time of the neutrino factory.
Furthermore, we discuss the uncertainty on the average Earth electron 
density along the neutrino path, and study its relevance in the range
$O(1$--$10\%)$.

In section 2, we analyse the sensitivity to CP violation in the 
full range $-180^\circ < \delta < 180^\circ$ and clarify the 
correlation of $\delta$ with the unknown parameter $\theta_{13}$. 
In section 3, the results of ref. \cite{golden} on the simultaneous
fits to $\theta_{13}$ and $\delta$ are extended to the full range of $\delta$ 
and $\theta_{13}<1^\circ$. Section 4 describes the method to include 
in the analysis the errors on the atmospheric and solar parameters and on 
the Earth matter density, taking properly into account the 
corresponding correlations between neutrinos and antineutrinos, as well 
as between different energy bins and baselines (when more than one baseline
is considered). 
The results of the fits including 
these errors and the discussion of the relative importance 
of each of them are then presented. 
Section 5 states our conclusions. 
 
\section{Sensitivity to CP violation} 

Given the high intensity expected at the neutrino factory,  
the effects of the solar mass difference, $\Delta m^2_{12}$, are not negligible 
over terrestrial distances in the LMA-MSW scenario, opening the 
way to observing leptonic CP violation. 
 
The best way to measure $\delta$ and $\tetaot$ is through the subleading 
transitions $\nu_e\rightarrow \nu_\mu$ and 
${\bar \nu}_e\rightarrow \bar{\nu}_\mu$. 
They can be measured at 
a neutrino factory by searching for wrong--sign muons \cite{geer,dgh} while running 
in both polarities of the beam, i.e. $\mu^+$ and $\mu^-$ respectively.
 
The exact oscillation probabilities in matter when no mass difference 
is neglected have been derived analytically in \cite{zs}. However, the 
physical implications of the formulae in \cite{zs} are not easily derived. 
Defining $\Delta_{ij} \equiv \frac{\Delta m^2_{ij}}{2 E}$, a convenient and 
precise approximation is obtained by expanding to second order 
in the following small parameters: 
$\tetaot$, $\Delta_{12}/\deltt$, $\Delta_{12}/A$ and $\Delta_{12} \, L$. 
The result is (details of the calculation can be found in \cite{golden}):
\bea
P_{\nu_ e \nu_\mu ( \bar \nu_e \bar \nu_\mu ) } & = & 
s_{23}^2 \sin^2 2 \tetaot \, \left ( \frac{ \delot }{ \tilde B_\mp } \right )^2
   \, \sin^2 \left( \frac{ \tilde B_\mp \, L}{2} \right) \, + \, 
c_{23}^2 \sin^2 2 \theta_{12} \, \left( \frac{ \Delta_{12} }{A} \right )^2 
   \, \sin^2 \left( \frac{A \, L}{2} \right ) \nn \\
& + & \label{approxprob}
\tilde J \; \frac{ \Delta_{12} }{A} \, \frac{ \delot }{ \tilde B_\mp } 
   \, \sin \left( \frac{ A L}{2}\right) 
   \, \sin \left( \frac{\tilde B_{\mp} L}{2}\right) 
   \, \cos \left( \pm \delta - \frac{ \delot \, L}{2} \right ) \, , 
\label{hastaelmogno}
\eea
where $L$ is the baseline, $\tilde B_\mp \equiv |A \mp \delot|$ and the 
matter parameter, $A$, is given in terms of the average electron 
number density, $n_e(L)$,  as $A \equiv \sqrt{2} \, G_F \, n_e(L)$, 
where the $L$-dependence will be taken from \cite{quigg}.
$\tilde J$ is defined as 
\be
 \tilde J \equiv \cos \theta_{13} \; \sin 2 \theta_{13}\; \sin 2 \theta_{23}\;
 \sin 2 \theta_{12}.
\ee

In the limit  $A\rightarrow 0$, 
this expression reduces to the simple formulae in vacuum 
\bea
P_{\nu_ e\nu_\mu ( \bar \nu_e \bar \nu_\mu ) } & = & 
s_{23}^2 \, \sin^2 2 \tetaot \, \sin^2 \left ( \frac{\delot \, L}{2} \right ) + 
c_{23}^2 \, \sin^2 2 \theta_{12} \, \sin^2 \left( \frac{ \Delta_{12} \, L}{2} \right ) \nn \\
& + & \tilde J \, \cos \left ( \pm \delta - \frac{ \delot \, L}{2} \right ) \;
\frac{ \Delta_{12} \, L}{2} \sin \left ( \frac{ \delot \, L}{2} \right ).  
\label{vacexpand} 
\eea
In the following we will denote by atmospheric, 
$P^{atm}_{\nu ( \bar \nu ) }$, solar, $P^{sol}$, and interference 
term, $P^{inter}_{\nu ( \bar \nu) }$, the three terms 
in eqs.~(\ref{hastaelmogno}).  

\begin{figure}[ht]
\begin{center}
\epsfig{file=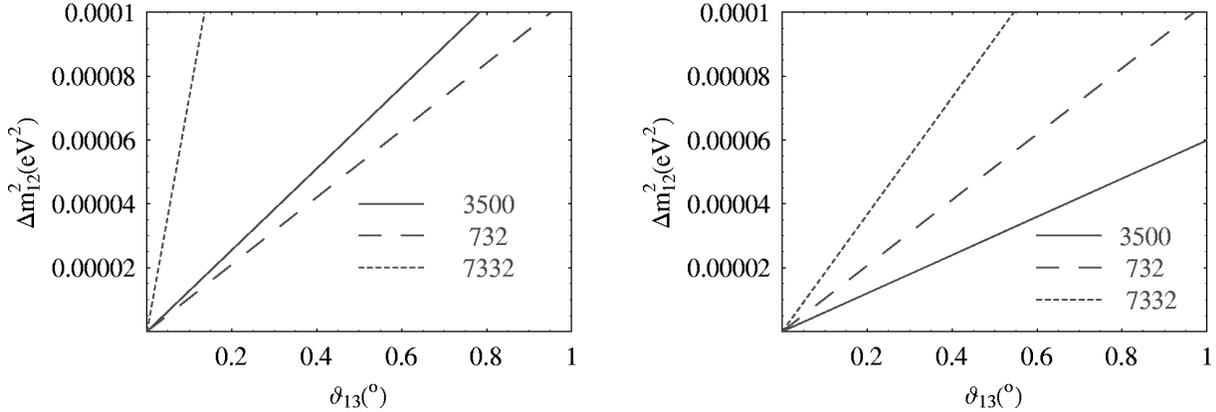,width=16cm}
\end{center}
\caption{\it Contours $P^{atm}_\nu = P^{sol}$ (left) and 
$P^{atm}_{\bar \nu} = P^{sol}$ (right)  on the plane $\theta_{13}, 
\Delta m_{12}^2$, for the three reference baselines. 
$\Delta m_{23}^2 = 3 \times 10^{-3}$ eV$^2$ and
$\theta_{12} = \theta_{23} = 45^\circ$. }
\label{regimes1}
\end{figure}

It is easy to show that
\bea
 |P^{inter}_{\nu(\bar \nu)}| \leq P^{atm}_{\nu ( \bar \nu ) } + P^{sol},
\eea
implying two very different regimes. When $\theta_{13}$ is 
relatively large or $\Delta m^2_{12}$ small, the probability is dominated
by the atmospheric term, since $P^{atm}_{\nu (\bar \nu )}\gg P^{sol}$. We will 
refer to this situation as the atmospheric regime. Conversely, when 
$\theta_{13}$ is very small or $\Delta m^2_{12}$ large, the solar 
term dominates $P^{sol} \gg P^{atm}_{\nu (\bar \nu )}$. This is the solar
regime. Fig.~\ref{regimes1} illustrates the separation between the two 
regimes on the plane $\Delta m^2_{12}$ and $\theta_{13}$ for neutrinos
and antineutrinos, as derived from eq.~(\ref{hastaelmogno}). The area to
the right(left) of the curves corresponds to the 
atmospheric(solar) regime. 

We discuss next the subtleties in the measurent of CP violation in both 
regimes. 

\subsection{Correlation between $\delta$ and $\theta_{13}$}

The oscillation probabilities of eq.~(\ref{hastaelmogno}), 
from whose measurement $\delta$ could be extracted, depend as well on 
$\theta_{23}, \Delta m^2_{23},\theta_{12}, \Delta m^2_{12}, A$ and 
$\theta_{13}$.
Uncertainties in the latter quantities can then hide the effect of 
CP violation. Although the first five of these parameters 
are expected to be known at the time 
of the neutrino factory with a good accuracy, $\theta_{13}$ 
might well remain unknown. It is essential then to understand whether the 
correlation between $\theta_{13}$ and $\delta$ can be resolved in such 
a way that CP violation is measurable.  
The effect of the uncertainties in the remaining parameters will be analysed in 
section 5. 
 
 Consider a single beam polarity and 
a fixed neutrino energy and baseline. 
The expansion of eq.~(\ref{hastaelmogno}) to second order in $\tetaot$ leads 
to 
\bea
P_{\nu_e \nu_\mu ( \bar \nu_e \bar \nu_\mu ) } =  
X_\pm
\theta^2_{13} \, + \, 
Y_\pm \tetaot
   \, \cos \left( \pm \delta - \frac{ \delot \, L}{2} \right ) + P^{sol} \, , 
\label{hastaelmogno2}
\eea
 with obvious assignations for the coefficients $X$ and $Y$, which are 
independent of  $\tetaot$ and $\delta$. Note that the solar term $P^{sol}$ is
the same for neutrinos and antineutrinos. 

Consider for instance $P_{\nu_e \nu_\mu}$. The question is how many 
values of  $(\tetaot, \delta)$ give the same probability than 
some central 
values chosen by nature $(\bar{\theta}_{13}, \bar{\delta})$.

 This requirement can be solved simply for $\theta_{13}$ as a function of 
$\delta$:
\bea
\tetaot & = & -\frac{Y_+}{2 X_+} \cos\left(\delta - \frac{ \delot \, L}{2} 
\right) \nonumber\\
& \pm & {\rm Sqrt}\left[\left(\frac{Y_+}{2 X_+} 
\cos\left( \delta - \frac{ \delot \, L}{2} \right)\right)^2 +
 \frac{1}{X_+} (P_{\nu_e\nu_\mu}(\bar{\theta}_{13},\bar{\delta})-P^{sol})\right].
\label{corr}
\eea
Eq.~(\ref{corr}) is a curve\footnote{The sign has to be chosen so that 
$\theta_{13} > 0$. Discontinuities in $\delta$ can arise only when
the argument of the square root becomes negative.} of equal probability on 
the plane $(\theta_{13}, \delta)$, which for most
of the parameter space spans the whole range of $\delta$. 
It follows that, at any baseline, it is 
not possible to determine $\delta$ from the measurement of wrong--sign
muons at fixed neutrino energy  with a single beam polarity. 

The analogous exercise for antineutrinos can be carried out 
 resulting in a different equal-probability curve, with the
substitutions in eq.~(\ref{corr}): 
$\delta \rightarrow -\delta$, $X_+(Y_+) \rightarrow X_-(Y_-)$. 
Assume now that both the neutrino and antineutrino oscillation 
probabilities have been measured, 
still at fixed (anti)neutrino energy and baseline. 
The question is if the two equal-probability curves intersect at values
of $(\tetaot, \delta)$ different from ($\bar{\theta}_{13}, \bar{\delta}$).
  This condition implies equating 
eq.~(\ref{corr}) to the corresponding one for antineutrinos 
and solving for $\delta$, for small $\tetaot > 0$. 
The resulting
equation is rather complicated, but simplifies 
considerably in the atmospheric 
and extreme solar regimes.

\subsubsection{Atmospheric regime}

In this regime it is safe to keep terms only up to first
order in $Y_+/X_+ (Y_-/X_-)$ in eq.~(\ref{corr}). As a result only the 
solution of eq.~(\ref{corr}) with $+$ sign in front of the square root 
is acceptable since $\theta_{13} > 0$. Eq.~(6) simplifies to:
\bea
\tetaot & = & \bar{\theta}_{13} -\frac{Y_+}{2 X_+} 
\left[\cos\left(\delta - \frac{ \delot \, L}{2}\right) 
- \cos\left(\bar{\delta} - \frac{ \delot \, L}{2}\right) \right].
\label{corratm}
\eea
The equation for $\delta$ is then obtained
from equating  eq.~(\ref{corratm}) for neutrinos to that for 
antineutrinos. The problem ammounts to finding 
the roots of a function of $\delta$ which is continuous and periodic. Since it
 must have at least one root at $\delta = \bar{\delta}$, 
by periodicity  there must be at 
least a second root in the range $-180^\circ < \delta < 180^\circ$. 

The second solution for $\delta$ in this approximation is:
\bea
\sin \delta - \sin \bar{\delta}& =&- 2  \, \frac {\sin \bar{\delta} - z \,\cos \bar{\delta}}
{1+ z^2},\nn\\
\cos \delta - \cos \bar{\delta}& =& 2 \, z \, \frac {\sin \bar{\delta} - z \,\cos \bar{\delta}}
{1+ z^2},
\label{cosdelta}
\eea
where
\be
z\equiv \frac{C_+}{C_-} \, \tan \frac{\Delta_{13} L}{2},\,\,\,;\,\,\,
C_\pm \equiv \frac{1} {2}( \frac{Y_+}{X_+}\pm \frac{Y_-}{X_-}).
\ee
The corresponding value of $\tetaot$ is:
\bea
\tetaot= {\bar{\theta}}_{13}
-\frac{1}{2} \;\frac{\sin {\bar\delta} - z \cos {\bar\delta}}{1+z^2} \,
\frac{C_+^2 - C_-^2}{C_-} \, \sin \frac {\Delta_{13}L}{2}.
\eea

 Only for the value of $\bar \delta$ satisfying 
\bea
\tan {\bar \delta} = z
\eea
do the two solutions degenerate into one. 
Except for this particular point, there are two degenerate solutions with the 
penalty that, if nature has been perverse in her choice of ${\bar \delta}$, one solution may correspond to CP-conservation and its image not, and viceversa.

\begin{figure}[ht]
\begin{center}
\epsfig{file=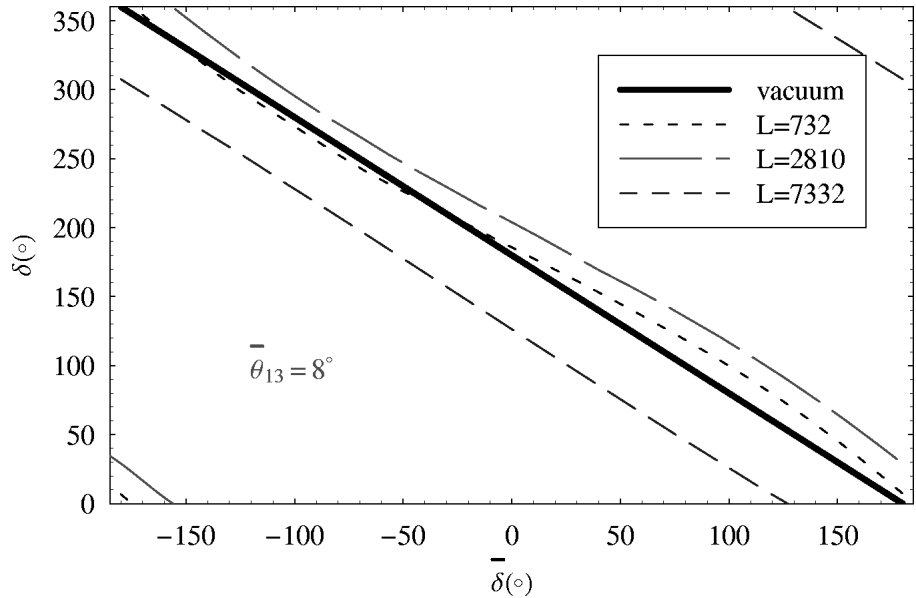, width=12cm}
\end{center}
\caption{\it Degenerate value 
of $\delta$ as a function of true value $\bar\delta$, 
for ${\bar\theta}_{13}=8^\circ$ and three different baselines; the
vacuum result $\delta= \pi -\bar{\delta}$ is also shown.}
\label{delta_atm}
\end{figure}

In vacuum this is not the case. Eq.~(\ref{cosdelta}) in the vacuum 
limit: $C_-\rightarrow 0$ or $z\rightarrow\infty$, 
gives $\delta= \pi -\bar{\delta}$ so that only 
for ${\bar \delta}=\pm \pi/2$ there is no degeneracy. 
Then the two solutions either break or conserve  CP.

In Fig.~(\ref{delta_atm}), we show the result 
of $\delta$ as a function of $\bar{\delta}$ for 
${\bar \theta}_{13} = 8^\circ$ for three reference baselines together 
with the vacuum result. The difference between $\delta$ and 
${\bar \delta}$ is maximal close to $\bar\delta = 0^\circ, 180^\circ$. 

It is interesting to consider the different impact of these degenerate
solutions at different baselines. At short baselines, the oscillation 
probabilities for neutrinos and antineutrinos are approximately the same for two reasons: 1) the relative size of the $\sin \delta$ versus $\cos \delta$ term 
in eq.~(\ref{hastaelmogno}) is $\tan(\Delta_{13} L/2) \ll 1$, 2) matter
effects are irrelevant with the solutions approaching the vacuum case 
\cite{golden,lipari}. Indeed, the expansion of eq.~(\ref{corratm}) for
 $\Delta_{13} L/2 \ll 1$ simplifies to
\bea
\tetaot & \simeq & \bar{\theta}_{13} -\frac{Y_+}{2 X_+} 
\left(\cos\delta -  \cos\bar{\delta}\right)\,. 
\label{corratms}
\eea
The same equation holds for antineutrinos, since $X_+(Y_+) = X_-(Y_-)$ 
in this approximation. The two equations have collapsed into one,
and consequently we expect to find a continuum curve of solutions
$(\theta_{13}, \delta)$ of the approximate form given by eq.~(\ref{corratms}).  
As the baseline increases the probabilities for neutrino and antineutrino
oscillations start to differ, not only due to the term in $\sin
\delta$, but also because of the matter effects. A shift in $\delta$
cannot in general be then compensated in the neutrino and antineutrino
probabilities by a common shift of $\theta_{13}$, and only the
two--fold degeneracy discussed above survives.

\subsubsection{Solar regime}
\label{sectsol}

In this regime the second term in eq.~({\ref{hastaelmogno}) dominates, although 
the first term cannot be neglected in the analysis of degenerate solutions even for very small 
values of $\bar{\theta}_{13}$. The reason is that there exist, at fixed neutrino energy
and baseline, a pair of values $(\tetaot, \delta)$ at which the first
and third terms in eq.~({\ref{hastaelmogno}) exactly compensate both for 
neutrinos and 
antineutrinos, in such a way that they are indistinguishable from 
the situation with ${\bar\theta}_{13}=0$ and any $\bar\delta$. 
It is easy to find 
these values by setting $\bar{\theta}_{13}=0$ in eq.~(\ref{corr}) and in the equivalent
equation for antineutrinos. $\delta$ is the solution of:
\bea
\tan \delta= - \frac{1}{z}\,,
\label{tetasol}
\eea
and the corresponding $\tetaot$ is: 
\bea
\tetaot = -\frac{Y_+}{X_+} \cos \left(\delta -\frac{\Delta_{13} L}{2}\right)\,.
\label{deltasol}
\eea
Taking as an example $\Delta m^2_{23} = 3 \times 10^{-3}$ eV$^2$, $L=2810$ km and 
$E_\nu = 0.3 E_\mu, E_\mu= 50$ GeV, this point is:
\bea
\tetaot \sim 1.5^\circ,\;\;\;\;\;\; \delta \sim -165^\circ.
\eea
As we will see in the next section, this solution is clearly seen in the 
fits. Alike to the pattern in the atmospheric regime, this degeneracy occurs
only at fixed neutrino energy and baseline.

In summary, this analysis implies that, even with the information from 
both beam polarities, there are in general two equally probable solutions,  
at fixed neutrino energy and baseline, for the parameters $\theta_{13}$ and 
$\delta$. This is of course applicable 
to any experiment that tries to measure simultaneously $\theta_{13}$ and 
$\delta$ from the subleading transitions $\nu_e\nu_\mu$ and ${\bar \nu}_e{\bar\nu}_\mu$.
These degeneracies can in principle be resolved by exploiting 
the energy and baseline dependence of the oscillation signals, 
as will be discussed later on. 
 We point out that a supplementary measurement of the channel 
$\nu_e\nu_\tau({\bar \nu}_e{\bar\nu}_\tau)$ could also be of great help in resolving 
the degeneracies, if performed with enough accuracy, as the 
$\delta$-dependent terms in the oscillation probabilities have the opposite sign 
than for the $\nu_e\nu_\mu({\bar \nu}_e{\bar\nu}_\mu)$ transitions. The 
experimental challenge of measuring those transitions is however much greater 
and we will 
not follow this avenue here.

\section{Simultaneous determination of $\delta$ and $\theta_{13}$}

In \cite{golden} a detailed study of the possibility to determine 
simultaneously $\theta_{13}$ and $\delta$ was performed, but only 
the range $\theta_{13} > 1^\circ$ was 
explored for the LMA-MSW solution. According to Fig.~\ref{regimes1}, this
corresponds to the atmospheric regime. 
However, the sensitivity to $\tetaot$ for the SMA-MSW solution 
(where only the term $P^{atm}$ survives) was found to extend to smaller
values: $\theta_{13} \sim 0.15^\circ$ at 
$\Delta m^2_{23} = 3\times 10^{-3}$ eV$^2$. For the LMA-MSW scenario, 
such a low value of $\tetaot$ lies 
well inside what we have named the solar regime, 
which is thus also relevant in the study of CP-violation for 
$\theta_{13} < 1^\circ$.

It is very illustrative to consider the 
different behaviour of the CP asymmetries, 
defined in \cite{cabibbo, dgh, dghr}, in both regimes, as an indication 
of the sensitivity to CP violation. 

The $\delta$-dependent terms in the oscillation probabilities 
of eq.~(\ref{hastaelmogno})  are linearly suppressed on the two small 
parameters: $\Delta m^2_{12}$  and $\theta_{13}$. 
In the atmospheric regime the leading term, $P^{atm}$, 
does not depend on $\Delta m^2_{12}$, while it is 
quadratically dependent on $\theta_{13}$. The sensitivity 
to CP violation decreases thus linearly with $\Delta m^2_{12}$ while 
it is rather stable as $\theta_{13}$ decreases \footnote{Although the 
relative importance of the terms in $\delta$ with respect to the leading term 
increases with decreasing $\theta_{13}$ \cite{romanino}, the statistical 
significance remains constant \cite{dghr}.}. There is of course a 
limit to this CP-insensitivity 
to $\theta_{13}$ and this is precisely when we enter the solar regime. 
In the solar regime the role of both parameters is interchanged and while 
the sensitivity to CP violation decreases linearly with $\theta_{13}$, it 
remains rather flat with $\Delta m^2_{12}$. 

This behaviour is shown in Fig.~\ref{regimes}, which displays 
the significance of the
 CP-odd asymmetries  defined in \cite{cabibbo, dgh, dghr} as a function of 
$\theta_{13}$, for $\Delta m^2_{12}$ fixed, and viceversa. In both
figures the change in behaviour coincides roughly with the   
limit between the atmospheric and solar regimes. 
These asymmetries have been obtained for a muon 
beam of $50\,\gev$ providing $10^{21}$ useful $\mu^+$ and $\mu^-$ decays and 
a 40 Kton magnetized iron detector \cite{lmd}, which is our 
working setup in the present work, as it was in \cite{golden}.

\begin{figure}[t]
\begin{center}
\epsfig{file=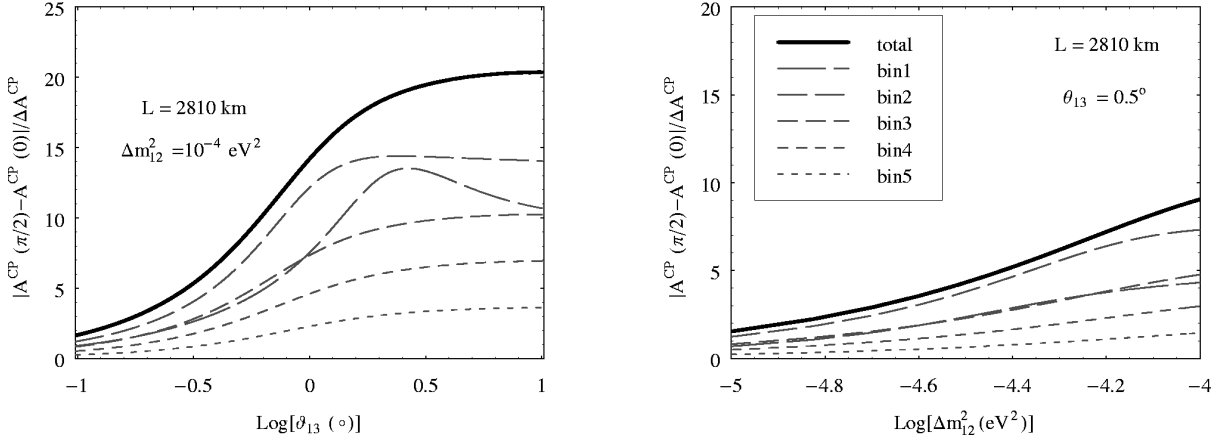, width=16cm}
\end{center}
\caption{\it Significance of the asymmetries at $L = 2810$ km as a function of 
$\theta_{13}$ for $E_\mu=50$ GeV, $\Delta m^2_{12} = 10^{-4}$ eV$^2$ 
(left figure) 
and $\Delta m^2_{12}$ for $\theta_{13} = 0.5^\circ$ (right figure). 
$\Delta m_{23}^2 = 3 \times 10^{-3}$ eV$^2$ and
$\theta_{12} = \theta_{23} = 45^\circ$. The five dashed curves correspond to 
neutrino energy bins of 10 GeV, and the solid curve is the average.}
\label{regimes}
\end{figure}

In \cite{golden} (see also \cite{bcr}) $\delta$ and $\theta_{13}$ 
were extracted from a spectral fit of wrong--sign muons signals
for both polarities, using the exact oscillation probabilities in the approximation of constant Earth matter density. Note that an analysis based on 
CP asymmetries alone does not use all the available information.  
The observables used were the number of 
wrong--sign muons in five bins of energy for both beam polarities:
\begin{equation}
N^\lambda_{i,\pm},
\end{equation}
where $\lambda$ labels the baseline, i the energy bin and $\pm$ 
the sign of the decaying muons. 
These numbers are given by:
\begin{equation}
N^\lambda_{i,\pm} = \int^{E_i +\Delta E}_{E_i} \; 
\Phi_{\nu(\bar{\nu})}(E_\nu,L) \; \sigma_{\nu(\bar{\nu})}(E_\nu) \;
 P_{\nu(\bar{\nu})}(E_\nu, L, \theta_{13}, \delta, {\alpha})   
\end{equation}
where ${\alpha}$ is the set of remaining oscillation parameters: $\theta_{23}, \theta_{12}, \Delta m^2_{23}, \Delta m^2_{12}$ and the matter parameter
$A$, which were taken as known. $\Phi_{\nu(\bar{\nu})}$ denote the neutrino 
fluxes and $\sigma_{\nu(\bar{\nu})}$ the DIS cross sections.

Simultaneous $\chi^2$ fits of the parameters $\delta$ and $\theta_{13}$ were
performed for three reference baselines $L=732$ km (the Cern to Gran Sasso 
distance or Fermilab to Soudan), $3500$ km (to be replaced in this work
 by $2810$ km, the distance from Cern to La Palma) and $7332$ km, 
as well as for various combinations of them. 

The $\chi^2$ at a fixed baseline is of the generic form:
\begin{equation}
\chi_\lambda^2 = \sum_{i,j} \sum_{p,p'} \; (n^\lambda_{i,p} - N^\lambda_{i,p}) C_{i,p:,j,p'}^{-1} (n^\lambda_{j,p'} - N^\lambda_{j,p'})\,,
\label{chi2}
\end{equation}
where $C$ is the $2 N_{bin} \times 2 N_{bin}$ covariance matrix. $n^\lambda_{i,p}$ are the simulated ``data'' obtained from a Gaussian or Poisson smearing  
including backgrounds and efficiencies (for more details we refer to \cite{golden}). In the combination of two baselines we have:
\begin{equation}
\chi_{\lambda\lambda'}^2 = \sum_{l,l'} \sum_{i,j} \sum_{p,p'} \; (n^l_{i,p} - N^l_{i,p}) C_{l,i,p:l',j,p'}^{-1} (n^{l'}_{j,p'} - N^{l'}_{j,p'})\,,
\label{chi2c}
\end{equation}
where $C$ is now a matrix of dimension $4 N_{bin} \times 4 N_{bin}$

As the errors on the ${\alpha}$ were neglected,  
$C$ contained only statistical errors, $\delta n^l_{i,p}$, which were assumed to 
be independent for different $i$, $p$ and $l$:
\begin{equation}
C_{l,i,p;l',j,p'} \equiv \delta_{ll'} \;\delta_{ij} \;\delta_{pp'} \;({\delta n}^l_{i,p})^2.
\end{equation}

In \cite{golden} only the restricted range $0 < \delta < 90^\circ$ 
and $\theta_{13} > 1^\circ$ was explored. For this reason the 
degeneracies appearing when the full range of $\delta$ is considered 
were missed. We have repeated the analysis by considering the full 
range of $\delta$ and 
$\theta_{13}$. In all the plots that follow realistic efficiencies and 
backgrounds have been included. We have checked
 that the degenerate images in the plots shown below appear at the points 
indicated by the analysis of the previous section.
The false images are somewhat softened, as different neutrino and antineutrino
 energies enter in the analysis, but still visible. We present fits  
only for $\Delta m^2_{23} > 0$. The opposite case
gives better results: for $\Delta m^2_{23} < 0$, the statistics for the signals
of positive and negative wrong--sign muons are  
closer (so that the difference is more neatly seen) 
because matter effects enhance in this case $P_{{\bar \nu}_e {\bar \nu}_\mu}$, 
compensating to a large extent the difference in the neutrino DIS 
cross sections  $\sigma _\nu \simeq 2 \sigma_{\bar \nu}$.

All the results shown below correspond to central values of the parameters 
in the 
LMA-MSW scenario: $\Delta m^2_{12} = 10^{-4}$ eV$^2$,  
$\Delta m_{23}^2 = 3 \times 10^{-3}$ eV$^2$ and
$\theta_{12} = \theta_{23} = 45^\circ$, except in Fig. \ref{excl}, where 
the full range of  $\Delta m^2_{12}$ is considered.

\begin{figure}[ht]
\begin{center}
\epsfig{file=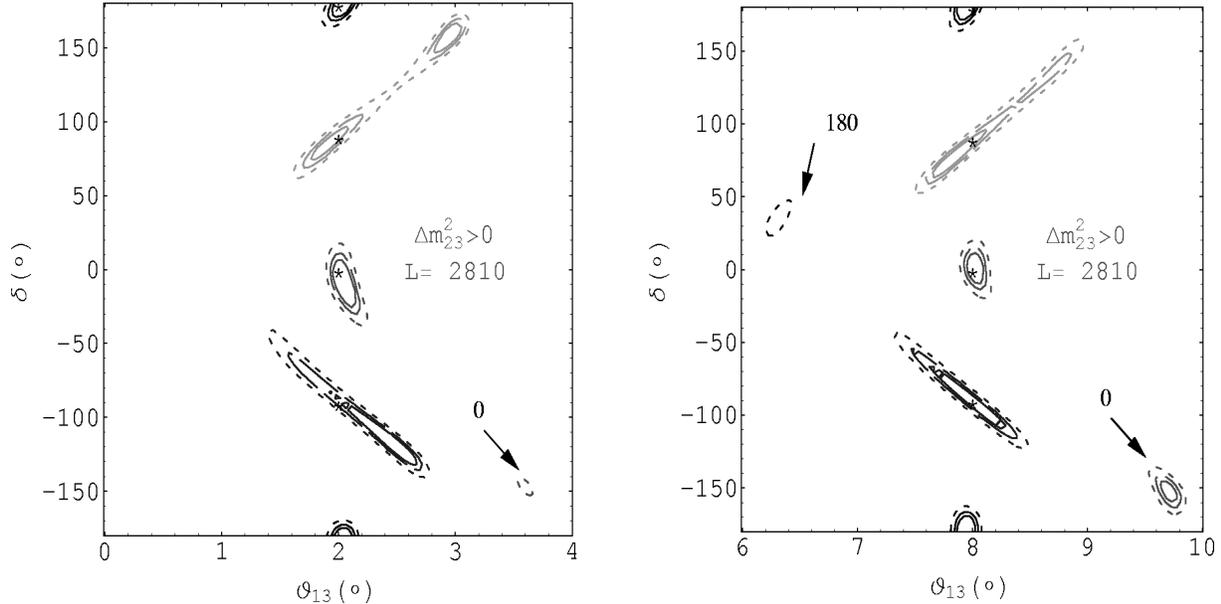, width=16cm,height=8cm} 
\end{center}
\caption{\it Simultaneous fits of $\delta$ and $\theta_{13}$ 
at $L = 2810$ km for different central 
values (indicated by the stars) of $\bar{\delta} 
= -90^\circ, 0^\circ, 90^\circ, 180^\circ$ and 
${\bar \theta}_{13}= 2^\circ$ (left), $8^\circ$ (right). The value 
of $\bar{\delta}$ for the degenerate solutions is also indicated. }
\label{golden1}
\end{figure}

\subsection{Atmospheric regime}

Recall that the pure CP-effects peak at distances in the range 
$2000$--$4000$ km \cite{ dgh, dghr,golden}.
In Figs.~\ref{golden1} we show the results of the fits 
including
efficiencies and backgrounds for $L=2810$ km for central values 
of ${\bar \delta} =- 90^\circ, 0^\circ, 90^\circ, 180^\circ$ and for 
$\bar{\theta}_{13}= 2^\circ$ (left) and $\bar{\theta}_{13}=8^\circ$ (right). 
The energy dependence of the signals is not 
significant enough (with our setup) to 
resolve the expected two-fold degeneracy. The second solution 
is clearly seen for the central value of ${\bar \delta}=0^\circ$ as an 
isolated island. 
For the central values ${\bar \delta} = -90^\circ$ and
${\bar \delta} = 90^\circ$, the degeneracy is responsible for the rather large 
contours which encompass the two solutions. Notice that as ${\bar \theta}_{13}$
diminishes the fake solution for $\bar{\delta} = 90^\circ$ moves towards 
$\delta = 180^\circ$, as it should (recall that, in the solar regime, the
vacuum fake image lies at $\delta=180^\circ$).

Figs.~\ref{golden2} show the fits for ${\bar \theta}_{13}=8^\circ$ at 
$L=732$ km and $7332$ km. In the former, the 
expected continous line of solutions of the form given 
by eq.~(\ref{corratms}) is clearly seen. The measurement of $\delta$ 
is thus impossible at this baseline if $\theta_{13}$ is unknown.  
In the longer baseline, 
the sensitivity to $\delta$ is similarly lost but for a different reason: the CP-signal is fading away (indeed the underlying degenerate solutions become much closer in $\theta_{13}$) and statistics is diminishing.

In Fig.~\ref{fig:ana2} we
show the result of combining any two baselines. The two-fold  degeneracy 
does not disappear completely in the combination of the two shorter baselines, 
while it does in the remaining two combinations. 
It is interesting  that, while
the shorter and longer baselines by themselves are not appropiate for CP 
studies, their combination is quite promising, due to the very different pattern
 shown by the correlation between $\tetaot$ and $\delta$ in each of them. 
 The overall conclusion 
is that the combination of any two distances is 
interesting regarding CP violation. 

\begin{figure}[ht]
\hskip -1.cm
\epsfig{file=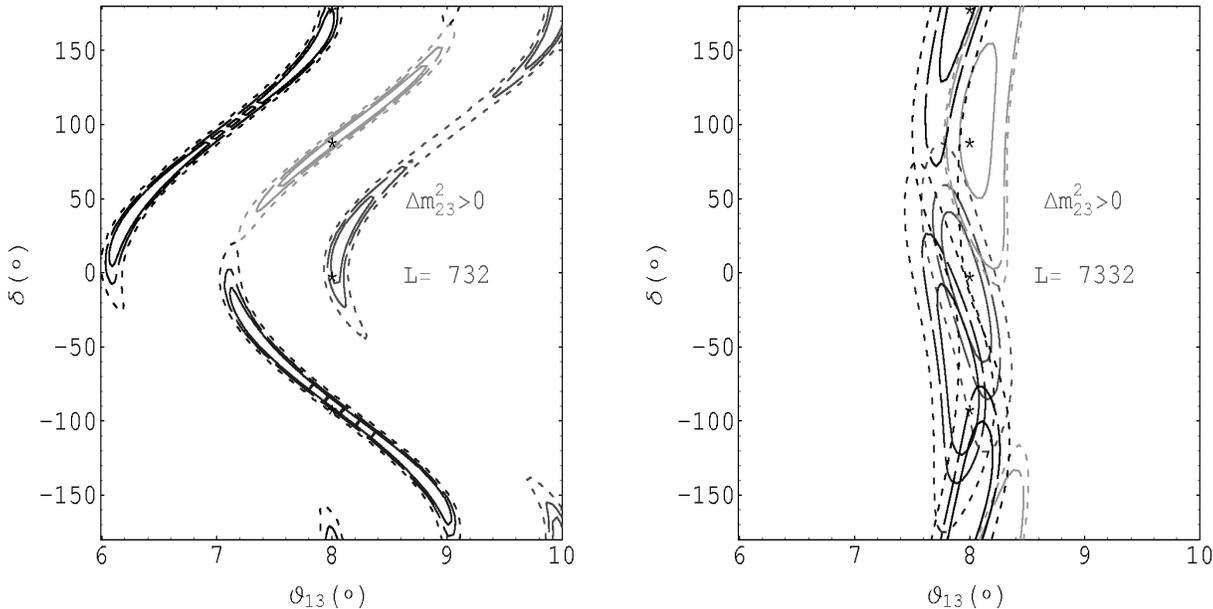, width=16cm,height=8cm} 
\caption{\it Simultaneous fits of $\delta$ and $\theta_{13}$ at $L = 732$ km (left) and $L= 7332$ km (right) for different central values of $\bar{\delta} = -90^\circ, 0^\circ, 90^\circ, 180^\circ$ and $\bar{\theta}_{13}= 8^\circ$. }
\label{golden2}
\end{figure}

\begin{figure}[ht]
\begin{center}
\mbox{
\epsfig{file=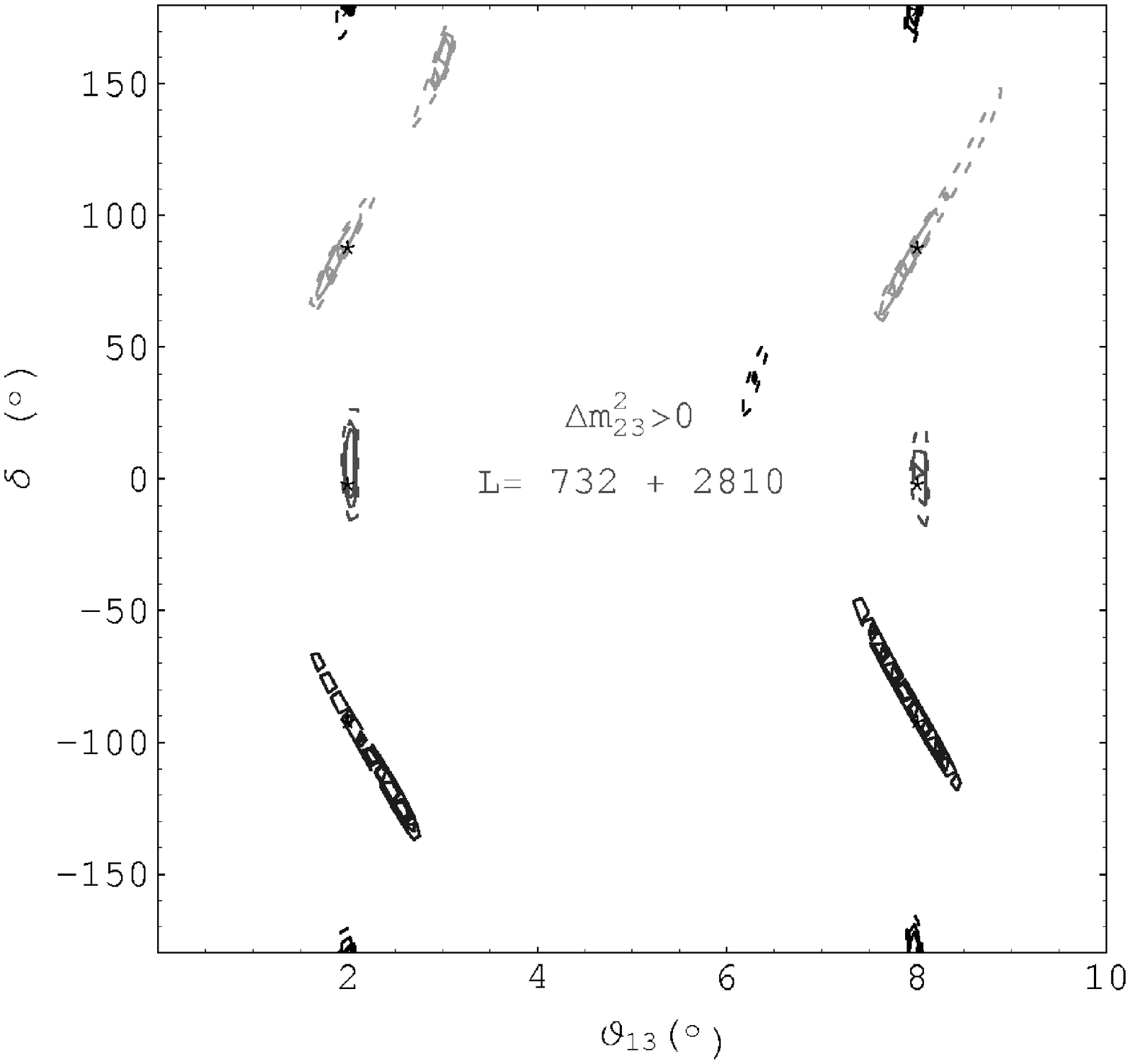,height=6cm,width=12cm}
}
\\
\mbox{
\epsfig{file=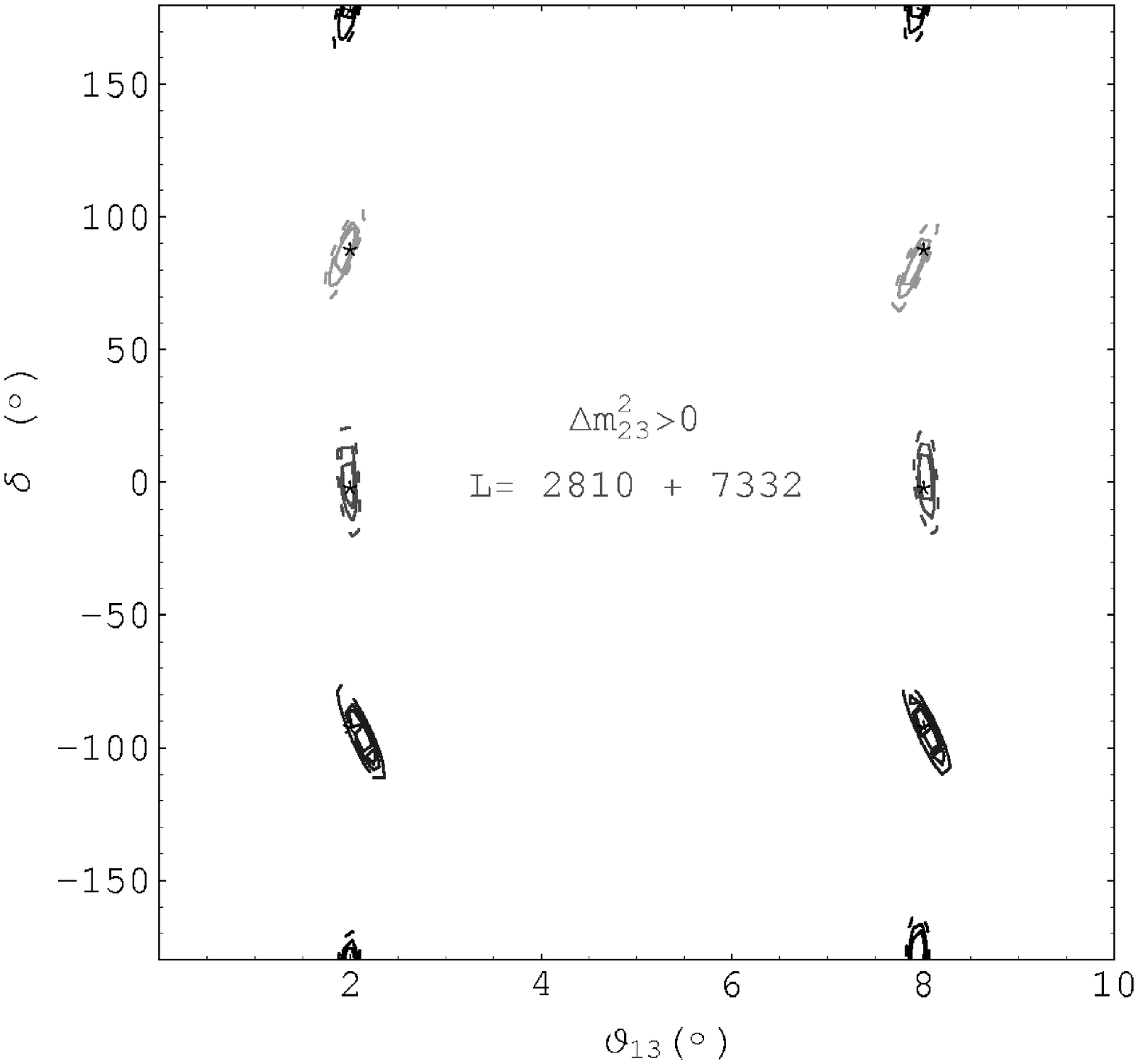,height=6cm,width=12cm}  
}    
\\
\mbox{
\epsfig{file=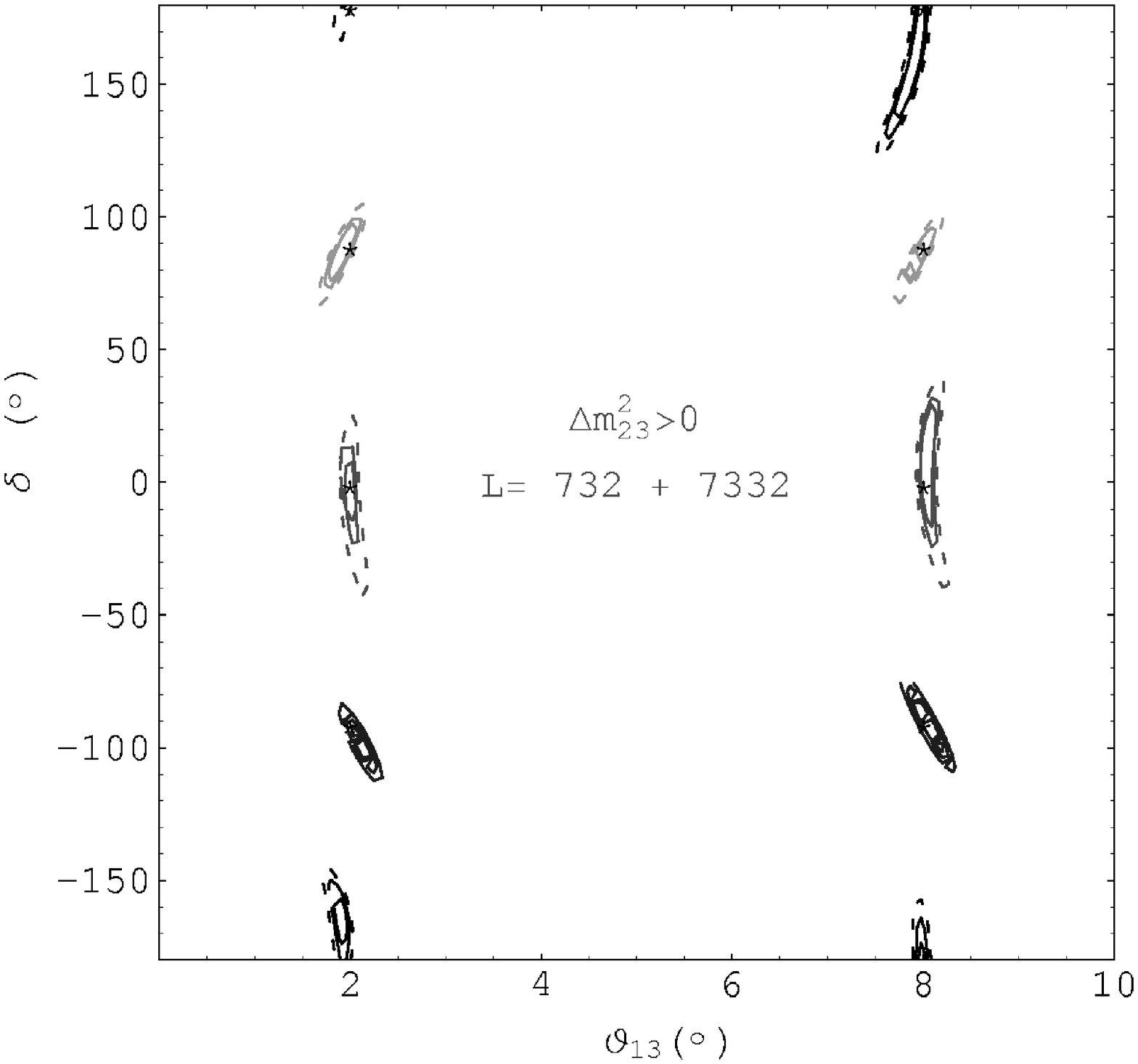,height=6cm,width=12cm}  
} 
\end{center}
\caption{Fits of $\delta$ and $\theta_{13}$ combining any two baselines.}
\label{fig:ana2}
\end{figure}

\subsection{Solar regime}

In Fig.~\ref{thpeq2810} we show the results of the fits 
including
efficiencies and backgrounds for $L=2810$ km for central values 
of ${\bar \delta} =- 90^\circ, 0^\circ, 90^\circ, 180^\circ$ and 
for ${\bar \theta}_{13}= 0.3^\circ$ (left) and ${\bar \theta}_{13}=0.6^\circ$ (right). 
Consider for instance the case ${\bar \theta}_{13}= 0.3^\circ$ (left): 
the degenerate images of the 
 four points chosen appear grouped at the right/lower side of the figure.
These are the solutions that mimic $\theta_{13} = 0$ as predicted in 
section \ref{sectsol}.  
The position of these solutions can be accurately predicted from the 
analysis  of the  previous section. 
The comparison of these figures with 
Fig.~\ref{golden1} illustrates the expected decrease of the
sensitivity to CP violation for very small $\theta_{13}$, consistent 
with  the behaviour of the significance of the CP-asymmetry in the solar 
regime shown in Fig.~\ref{regimes}. Note that at ${\bar \theta}_{13} = 0.3$, 
the sensitivity to CP is already lost for ${\bar \delta} = -90^\circ$.

\begin{figure}[ht]
\epsfig{file=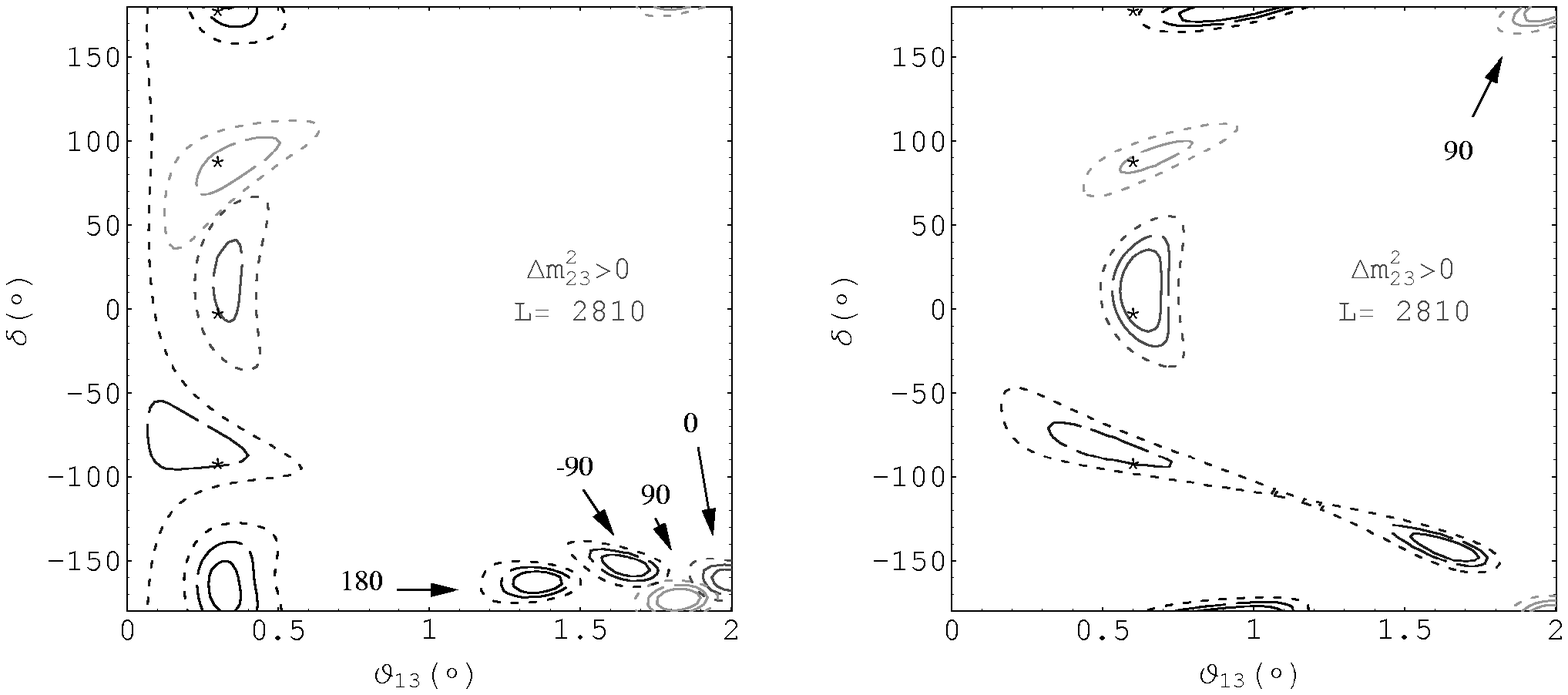, width=16.cm,height=8cm} 
\caption{\it Simultaneous fits of $\delta$ and $\theta_{13}$ at  
$L = 2810$ km, for different central values of $\bar{\delta} = -90^\circ,0^\circ,90^\circ,180^\circ$ and ${\bar \theta}_{13}= 0.3^\circ$ (left), $0.6^\circ$ (right). The value of $\bar{\delta}$ for the degenerate solutions is indicated. } 
\label{thpeq2810}
\end{figure}

In Fig.~\ref{thpeqsm} we
show the result of combining the baselines $L=2810$ km and $L = 732$ km with 
 the longer one. As in the atmospheric regime, the degeneracies are
nicely resolved in the combination of the intermediate and long baselines. 
However, in this case the combination of the short and long baselines
is not as good. Note the degenerate solution that survives 
for $\bar{\delta} = 90^\circ$ which appears centered at $\delta = - 90^\circ$. 
This is the result of the intersection of the approximately vertical 
contours found at $L=7332$ km with the $\cos \delta$ contours at 
$L=732$ km (see Fig. \ref{golden2}).

\begin{figure}[ht]
\epsfig{file=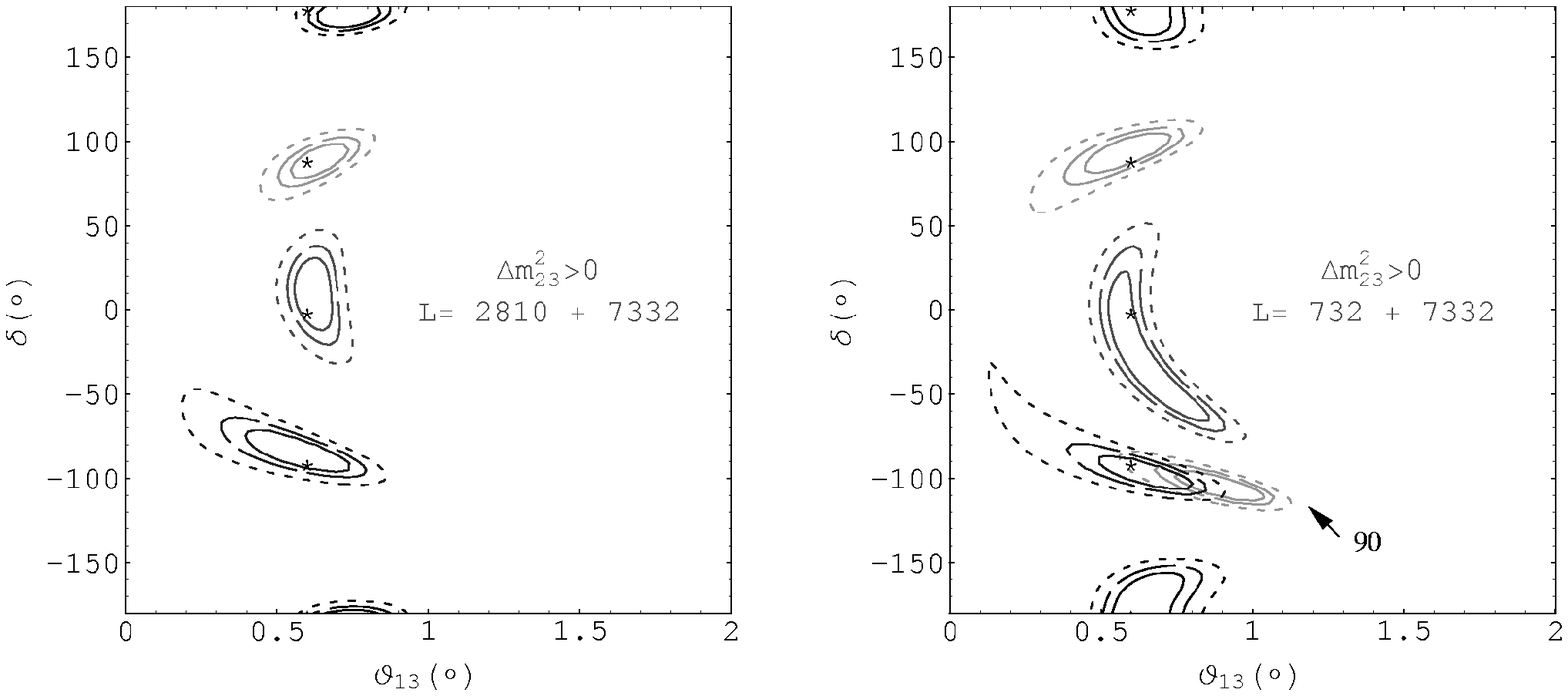, width=16.cm,height=8cm} 
\caption{\it Simultaneous fits of $\delta$ and $\theta_{13}$ at two 
combinations of baselines, $L = 2810 + 7332$ km (left) and
 $L = 732 + 7332$ km (right), for  
${\bar \theta}_{13}= 0.6^\circ$ and different central values 
of $\bar{\delta} = -90^\circ,0^\circ,90^\circ,180^\circ$.} 
\label{thpeqsm}
\end{figure}

\section{New analysis: inclusion of expected errors on oscillation parameters 
and matter density}

Consider the uncertainty in  
the theoretical parameters ${\alpha}$. These errors induce in general 
a correlation between the observables, $N^\lambda_{i,\pm}$, which has to be taken into account. The matrix $C$ of eq.~(\ref{chi2c}) is of the form:
\begin{equation}
C_{l,i,p;l',j,p'} \equiv \delta_{ll'}\; \delta_{ij} \;\delta_{pp'} \; ({\delta n}^l_{i,p})^2 
+ \sum_\alpha \; \frac{\partial N^l_{i,p}}{\partial \alpha} \frac{\partial N^{l'}_{j,p'}}{\partial \alpha}\;\sigma^2(\alpha),
\label{cm}
\end{equation}
where $\sigma(\alpha)$ is the 1$\sigma$ error on the parameter $\alpha$. 

 Recent analysis of the expected uncertainty in the knowledge of the 
atmospheric parameters at the neutrino factory 
 indicate a $\sim 1\%$ uncertainty in  $\Delta m_{23}^2$ and 
$\sin^2 2 \theta_{23}$ 
\cite{barger,concha}\footnote{ Although these analyses have 
been done for the SMA-MSW solution or assuming that the solar parameters
are known, we will assume that in the LMA-MSW scenario the errors on the 
solar parameters or in the matter term do not change this result.} 
For the solar parameters we include the results of the 
analyses of the Kamland reach \cite{kamland}: $2\%$ error in 
$\Delta m_{12}^2$ and 
$\pm 0.04$ in $\sin^2 2\theta_{12}$, for maximal $\theta_{12}$, both 
at $1\sigma$. 
Slightly smaller errors could be obtained for the product 
$(\sin 2 \theta_{12} \,\Delta m^2_{12})^2$ which is the 
combination entering the relevant oscillation 
probabilities \cite{privcom}, a refinement we neglect here. 
For the uncertainty on the matter parameter, $A$, we could not find 
any estimate in the literature.
The dispersion of the different models of the Earth density 
profile \cite{bahcall} indicates an uncertainty of $1$--$2\%$ for 
trajectories which 
do not cross the core, though. 
We consider a range between $1$--$10\%$ for illustration.

 The most important changes result from the uncertainty in $\theta_{23}$ and
in the matter parameter $A$, with the former affecting mainly the measurement 
of $\theta_{13}$ and the latter the sensitivity to $\delta$.

Recall Fig.~4, where no errors on the oscillation and matter parameters 
were included.
Fig.~\ref{allerrors_2810} (left) depicts the results 
for ${\bar \delta} = 90^\circ$ and $-90^\circ$ at $L=2810$ km, 
 including all errors 
(with an error in the matter parameter of
$1\%$) compared (right) with the situation in which only the error 
on the atmospheric angle $\theta_{23}$ is included. 
The two graphics in this figure are almost identical, showing that the dominant 
error is that of $\theta_{23}$. It affects mainly the determination
of $\theta_{13}$, a fact easy to understand: while 
the measurement of the leading transition $\nu_\mu\rightarrow \nu_\tau$ is 
sensitive to $\sin^2 2 \theta_{23}$, and this can be measured with a 
$1\%$ uncertainty,  the subleading transition 
$\nu_e\rightarrow \nu_\mu$ is proportional to $\sin^2 \theta_{23}$. For 
maximal $\theta_{23}$ mixing a $1\%$ relative error in the former 
translates into a $6\%$ in the latter. This is then the largest relative error 
of all the parameters that enter $P^{atm}$, which is dominant in most
of the parameter space. 
Note that the effect of the errors is more important for larger 
$\bar\theta_{13}$. 
In Fig.~\ref{allerrors_ml} we show the results for the best combination 
of baselines when all errors have been included. The resolution of the
degeneracies discussed in the previous sections is still achieved, but
the contours have become sizeably larger. 

\begin{figure}[ht]
\begin{center}
\epsfig{file=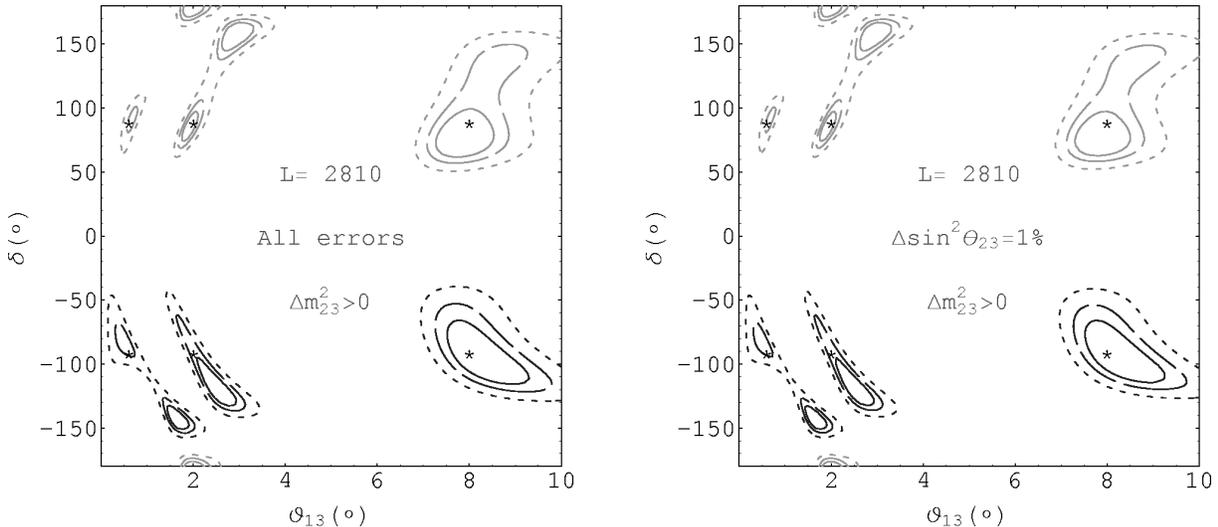, width=16cm} 
\end{center}
\caption{\it Fits of $\delta$ and $\theta_{13}$ for various central 
values of ${\bar \delta}$ and ${\bar \theta}_{13}$ at $L=2810$ km including 
all the errors  on the remaining parameters (left plot) with $\Delta A/A = 1\%$ and including only the error on $\theta_{23}$ (right plot).}  
\label{allerrors_2810}
\end{figure}

\begin{figure}[ht]
\begin{center}
\epsfig{file=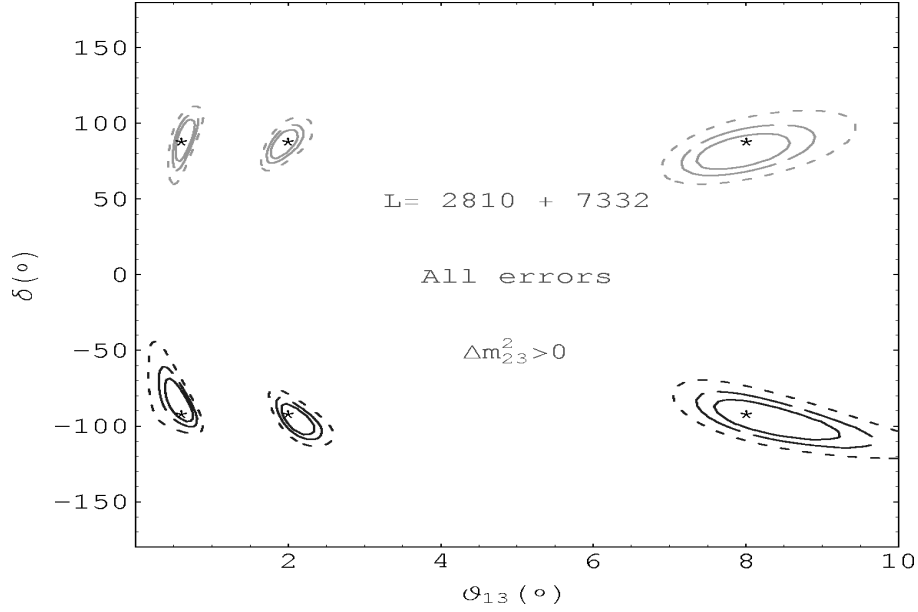, width=12cm,height=8cm} 
\end{center}
\caption{\it Fits of $\delta$ and $\theta_{13}$ combining the two baselines: $2810$ km and $7332$ km, for various central 
values of ${\bar \delta}$ and ${\bar \theta}_{13}$ including 
all the errors  on the remaining parameters with $\Delta A/A = 1\%$.}
\label{allerrors_ml}
\end{figure}

 The uncertainty in $A$ is more relevant for the determination of 
$\delta$, although if this error is controlled at the percent level 
the effect is negligible. 
In general, even for $\Delta A/A= 10\%$, the effect is  
far less important than the error induced by correlations 
between $\theta_{13}$ and $\delta$. This can be seen with 
just one baseline, where the degeneracies survive.
As an illustration, 
in Fig.~\ref{Aerror_2810}  for L=$2810$ km, $\Delta A/A$ is 
varied in the range $1-10\%$, with 
 $\bar{\delta}=90^\circ$, $\bar{\theta}_{13} =8^\circ$ 
 and the errors on the remaining oscillation parameters 
included.
 This is to be compared with 
Fig.~\ref{golden1} (right) where no uncertainties were 
assumed. The error in $\delta$ is seen to be mostly dominated
by the correlation of $\theta_{13}$ and $\delta$. 
For the combination $L=2810+7332$ km, where the degeneracies are resolved, 
 the effect of the error in A is more important in 
relative terms (compare Fig.~\ref{allerrors_ml} with the middle plot in Fig.~\ref{fig:ana2}): a $10\%$ error in $A$ is a $50\%$ 
increase of the error in $\delta$.
\begin{figure}[ht]
\begin{center}
\epsfig{file=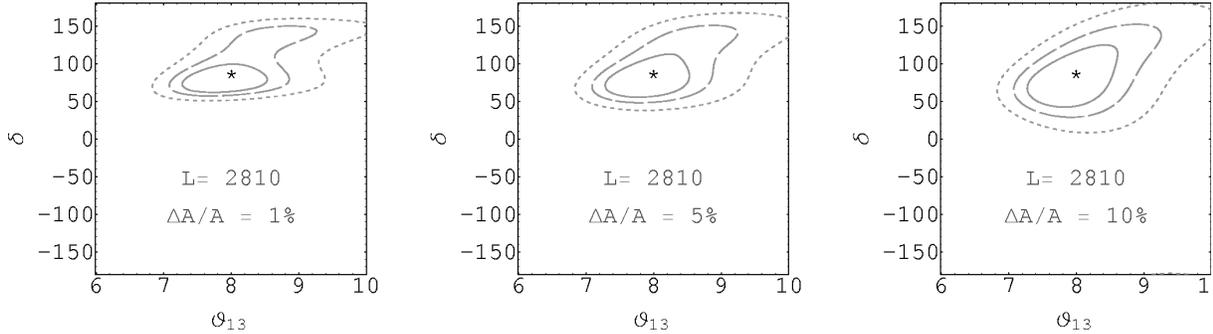, width=16cm}
\end{center}
\caption{\it 
Results from a simultaneous fits to $\theta_{13}, \delta$ for ${\bar \delta} = 90^\circ$ including errors on the oscillation parameters and varying the 
error on the average Earth matter density from $1\%=10\%$.}
\label{Aerror_2810}
\end{figure}
 It is clearly desirable   
to have a per cent control over the average Earth matter density, which 
does not look a priori unrealistic.

Recently, the authors of \cite{sato} have presented an analysis of the 
sensitivity to CP violation, including  
the errors on the oscillation and matter parameters, 
with quite different conclusions.
 In particular, they 
state that CP violation can only be measured in a small window at the shorter
baselines. There are a number of differences between their analysis and ours.
They do not consider the simultaneous determination
 of $\theta_{13}$ and $\delta$, 
but include and ad hoc $10\%$ error on $\theta_{13}$.  
They assume as well larger errors on the remaining solar and 
atmospheric parameters: a democratic $10\%$. 
We have seen that estimates from Kamland and the 
expectation from disappearance measurements at the neutrino factory
 give more optimistic
results. Finally, they do not take into account correlations of the  
errors that these parameters induce on the different observables: wrong--sign 
muons in different energy bins and different polarities. Neither they 
include experimental background and efficiencies.

Let us turn now to the case of smaller values of  $\Delta m^2_{12}$ allowed 
in the LMA-MSW range. For fixed 
$\theta_{13}$, the sensitivity
to $\delta$ decreases linearly with $\Delta m^2_{12}$ in the atmospheric regime
and more slowly in the solar one. 
For the plots in the atmospheric regime of this section  it does not necessarily imply, though, a linear scaling 
(with $\Delta m^2_{12}$) of the error in $\delta$, as the latter 
 is mostly dominated by the existence of degenerate solutions on the 
plane ($\theta_{13}, \delta$), whose
separation dos not follow such a linear pattern.

It is interesting to understand how much of the LMA-MSW range  
can be covered in the discovery of CP violation. This is illustrated in 
Fig.~\ref{excl} with a rough exclusion plot.  For the hypothetical nature values ${\bar \delta} = 90^\circ$  and
the best combination of 
baselines, $L=2810+7332$ km, the line corresponds to the 
minimum value of $\Delta m^2_{12}$ at which 
the $99\%$CL error on the phase reaches $90^\circ$ degrees, and is thus 
indistinguishable from $0^\circ$  or $180^\circ$ (i.e. no CP violation). 
The error 
on the phase is computed by taking the longest vertical size (upwards or
downwards, whichever is longest) of the
$99\%$CL contour from $90^\circ$. 
All errors on the parameters have been included. With this definition, there is sensitivity to CP violation for $\tetaot>$ few tenths of degree and 
$\Delta m^2_{12}> 3\times 10^{-4}$ eV$^2$.

An analogous plot in \cite{golden} indicated better sensitivity.
Several reasons account for the difference. First, only the possibility 
of separating $90^\circ$  
from $0^\circ$ (instead of  $180^\circ$, which is more 
constraining \cite{sato}) was considered there. Second,
the correlation between $\theta_{13}$ and $\delta$ was not taken into account: 
in order words, $\theta_{13}$ was fixed at its
true value ${\bar \theta}_{13}$ and only then the error on $\delta$ considered.
Finally, the errors on the remaining oscillation parameters and $A$ were
not included, although this makes a very small difference if $\Delta A/A = 1\%$.

\begin{figure}[ht]
\begin{center}
\epsfig{file=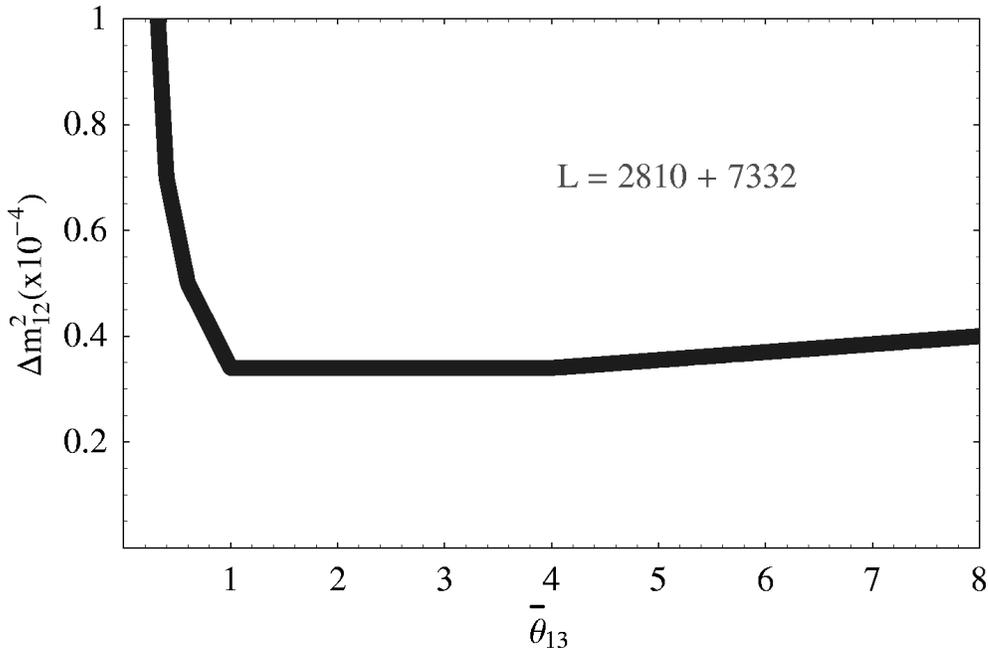, width=13cm}
\end{center}
\caption{\it 
Sensitivity reach for CP violation as defined in the text on the plane
$(\Delta m^2_{12}, {\bar \theta}_{13})$ for the combination of baselines
 L = 2810 and 7332 km. All errors are included. }
\label{excl}
\end{figure}

A final comment. As stated before, the problem of the correlations between 
$\tetaot$ and $\delta$ has to be faced by any experiment measuring 
just the $\nu_e\leftrightarrow\nu_\mu$ and $\bar\nu_e\leftrightarrow\bar\nu_\mu$ 
transitions. In particular this applies to the so-called 
``superbeams'' \cite{superbeams}: intense neutrino beams from 
pion (and kaon) decay. They could provide, though,  
very useful complementary information to the neutrino factory in 
disentangling $\theta_{13}$ and $\delta$, for their expected reach
$\theta_{13} > 3^\circ$\cite{superbeams}.

\section{Conclusions}

 A neutrino factory from muon storage rings, with muon energies of a few
 dozen GeV, is an appropiate facility to discover leptonic CP violation
 through wrong--sign muon searches.
 This requires that the solution to the neutrino solar deficit is confirmed 
to lie in the LMA-MSW regime, 
and the angle $\theta_{13}$ is larger than a few tenths of degree. 
Within this range, the sensitivity to CP-violation is lost only 
for the smaller values of the solar mass difference allowed by the 
LMA-MSW scenario. 

 At the hypothetical time of the neutrino factory, the value of 
the parameters $\tetaot$ and $\delta$ may be still unknown and will have to be 
simultaneously measured.
In this paper we have considered the full range of possible values of $\delta$.
 A relevant problem unearthed is the generic existence, at a given
(anti)neutrino energy and fixed baseline, 
of a second value of the set ($\tetaot,\delta$) which gives the same 
oscillation probabilities for neutrinos and antineutrinos than the 
true value chosen by nature. It is a generic challenge for any future facility.
The spectral analysis and the combination of baselines satisfactorily 
resolves this degeneracy.

 Furthermore, we have included in the analysis the expected uncertainty on 
the knowledge of the rest of the oscillation parameters ($\sin^2 \theta_{23},\,
\Delta m^2_{23},\,\sin^2\theta_{12},\,\Delta m^2_{12}$) and on the Earth 
electron density. Noticeable changes result from the error 
on $\theta_{23}$, which affects mainly the uncertainty 
in $\theta_{13}$, and from the 
uncertainty on the Earth matter profile, which affects mainly the extraction of
$\delta$. The latter uncertainty is of little consequence if at the level of a
few percent.
It seems pertinent to us that a detailed geological analysis of 
the planned baselines is performed, to reinforce the expectations as regards 
CP violation. 

Realistic background detection errors and experimental efficiencies have been 
included in the analysis.
The overall conclusion is that the optimal distance for studying CP-violation 
effects with neutrino energies of few dozens of GeV is still of $O(3000)$ km,
although the combination of two baselines, one of which being preferably
 a very long one, is very important in resolving degeneracies.

\section{Acknowledgements}

 We thank J.~Bahcall, A.~Cervera, A.~Donini, A.~de~Gouvea, P.~Lipari, H.~Murayama, A.~Pierce and S.~Rigolin  for useful
 conversations and exchanges. We are specially indebted to A.~Cervera, A.~Donini and S.~Rigolin for earlier collaboration on this subject and their 
contribution to the early stages of this work.
We acknowledge as well discussions 
with J. Sato. This work has been partially supported by CICYT projects
 AEN/97/1678, AEN/99/0692 and GCPA640 as well as the Generalitat Valenciana
project GV99-3-1-01.

\end{document}